\documentstyle[sprocl,citepunct]{article}
\input epsf.tex

\newcommand{\ev}{\text{eV}}

\newcommand{\gev}{\text{GeV}}
\newcommand{\tev}{\text{TeV}}

\newcommand{\be}{\begin{equation}}
\newcommand{\ee}{\end{equation}}

\newcommand{\ibid}{{\em ibid.}}
\newcommand{\eqref}[1]{Eq.~(\ref{#1})}

\newcommand{\rem}[1]{#1} % For arxiv
\newcommand{\agt}{ \mathop{}_{\textstyle \sim}^{\textstyle >} }
\newcommand{\alt}{ \mathop{}_{\textstyle \sim}^{\textstyle <} }
\newcommand{\text}[1]{ {\rm #1} }

\begin{document}

\title{THEORETICAL OVERVIEW: \\MOTIVATIONS FOR LEPTON FLAVOR VIOLATION
\rem{\footnote{Invited talk presented at New Initiatives in
Lepton Flavor Violation and Neutrino Oscillations with Very Intense
Muon and Neutrino Sources, 2-6 October 2000, University of Hawaii,
USA.}}
}

\author{JONATHAN L. FENG
\rem{\footnote{Electronic address: jlf@mit.edu}}
}

\address{Center for Theoretical Physics, Massachusetts Institute
of Technology\\ Cambridge, MA 02139 USA}

\maketitle\abstracts{In the coming years, experiments underway will
increase the sensitivity to charged lepton flavor violation by four
orders of magnitude.  These experiments will stringently probe weak
scale physics. I review the status of global symmetries in the
standard model and present several well-motivated models that predict
observable lepton flavor violation.  Finally, I describe what we might
learn from future experimental results, whether positive or null.}

\rem{\vskip-2.68in
\noindent
MIT--CTP--3066 \hfill hep-ph/0101122
\vskip2.42in
}

\section{Introduction}
\label{introduction}

The diverse topics discussed at this conference, ranging from charged
lepton flavor violation (LFV) to neutrino oscillations to proton
decay, have been brought together by a common experimental dependence
on high intensity muon and neutrino sources.  {}From a theoretical
perspective, however, these topics also have a unifying theme --- they
are all probes of global symmetries.  The exploration of symmetries
and symmetry breaking is, of course, central to particle physics.
These experiments form a comprehensive program to probe the standard
model's (broken) global symmetries, just as high energy colliders aim
to study thoroughly the standard model's (broken) local symmetries.

Global symmetries are brittle.  It is therefore somewhat surprising
that global symmetry breaking is currently observed only in neutrino
oscillations.  In the near future, however, experiments will increase
the sensitivity to LFV by four or more orders of magnitude.  These
experiments have the potential to uncover many exotic phenomena, which
may exist at a variety of scales~\cite{Kuno:1999jp}.  What is even
more compelling, however, at least to me, is that these experiments
will for the first time stringently probe the weak scale, where we
have strong reasons to expect new physics to appear.  There is thus
the promise of a fascinating interplay between high precision LFV
experiments and high energy experiments probing electroweak symmetry
breaking.  For this reason, results from upcoming LFV experiments,
whether positive or null, will have interesting implications.

I begin with a brief review of the standard model's global symmetries
in Sec.~\ref{sec:global}. Section~\ref{sec:mass} contains a simple but
telling model-independent analysis of the reach of present and future
experiments.  {}From this it will be clear that future LFV experiments
will be strong probes of new physics at the weak scale.  I will then
focus on some broad possibilities related to the physics of
electroweak symmetry breaking in Sec.~\ref{sec:EWSB}, and consider the
example of supersymmetry in Sec.~\ref{sec:supersymmetry}.  In
Sec.~\ref{sec:what}, I conclude with several illustrations of what we
might learn from future results.

\section{Global Symmetries of the Standard Model}
\label{sec:global}

The fields of the standard model, their spins, and their quantum
numbers under the three standard model gauge symmetries are:
\begin{equation}
\begin{array}{ccccccc}
         & \quad q \quad & \quad u \quad & \quad d \quad 
& \quad l \quad & \quad e \quad & \quad h \quad \rule[-1mm]{0mm}{5mm}\\
\quad \text{Spin}\ S \quad & \frac{1}{2} & \frac{1}{2} & \frac{1}{2} 
& \frac{1}{2} & \frac{1}{2} & 0 \rule[-1mm]{0mm}{5mm}\\
\text{SU(3)}  & 3 & 3 & 3 & 1 & 1 & 1 \rule[-1mm]{0mm}{5mm}\\
\text{SU(2)}  & 2 & 1 & 1 & 2 & 1 & 2 \rule[-1mm]{0mm}{5mm}\\
\text{U(1)}   & \frac{1}{6} & \frac{2}{3} & -\frac{1}{3} & -\frac{1}{2} 
& -1 & \frac{1}{2} \rule[-1mm]{0mm}{5mm}
\end{array}
\end{equation}
Renormalizability and gauge invariance restrict the interactions among
these fields to the form
\begin{equation}
{\cal L}_{\text{Yukawa}} = {y_e}_{ij} h \bar{l}_i e_j + {y_d}_{ij}
h \bar{q}_i d_j + {y_u}_{ij} h^c \bar{q}_i u_j \ ,
\label{Yukawa}
\end{equation}
where $i$ and $j$ are generational indices.  Through unitary rotations
in generation space, we may set ${y_e}_{ij} = \text{diag} (y_e,
y_{\mu}, y_{\tau})$.  The Yukawa couplings, then, conserve baryon
number $B$, total lepton number $L$, and the lepton flavor numbers
$L_e$, $L_{\mu}$, and $L_{\tau}$.  These are the global symmetries of
the standard model.

Global symmetries are broken in many possible ways, however.
Beginning with the most robust theoretically, even in the standard
model itself, non-perturbative quantum effects~\cite{'tHooft} break
most of the global symmetries, preserving only $B-L$.  Gravitational
interactions are also expected to break all global symmetries through,
for example, processes involving black holes. Both of these effects
are typically extremely suppressed, but demonstrate that global
symmetries are by no means sacred.  More experimentally relevant
examples arise in attempts to extend the standard model.  In grand
unified theories, for example, additional gauge bosons lead to global
symmetry violation: in SU(5) only $B-L$ is preserved, and in SO(10),
even this is broken.  Finally, as we will see below, attempts to
explain electroweak symmetry breaking often predict some form of
global symmetry violation.  In many cases, these predictions are
generically at levels far beyond current constraints.  It is a triumph
of the standard model, and a source of frustration for those
attempting to extend it, that its `accidental' global symmetries are
so accurately obeyed in nature.

\section{Mass Scales and Experimental Probes}
\label{sec:mass}

What is the current experimental status of global symmetries?  For a
model-independent answer, we turn to the formalism of effective field
theory.  If we expect the standard model to be valid up to some energy
scale $M$, the effects of physics above this scale may be accounted
for by shifts in the standard model couplings and a series of
non-renormalizable terms suppressed by powers of
$M$~\cite{Appelquist:1975tg}.  The terms of \eqref{Yukawa} are then to
be viewed as merely the first in a series.  At higher orders in $1/M$,
one finds non-renormalizable operators that violate the global
symmetries:
\begin{eqnarray}
B \text{-violating:} &&\ \ \frac{q q q l}{M^2}\ , \ \ldots 
\label{Bviolating} \\
L \text{-violating:} &&\ \ \frac{hhll}{M}\ , \ \ldots 
\label{Lviolating} \\
L_i \text{-violating:} &&\ \ \frac{h \; \bar{l} \sigma^{\alpha \beta} 
e \; F_{\alpha \beta}}{M^2} \ , \ 
\frac{\bar{l} l \bar{e} e}{M^2} \ , \ \ldots
\ ,
\label{Liviolating} 
\end{eqnarray}
where generational indices have been omitted.  The experimental status
of global symmetries may be roughly summarized by determining what
scales $M$ are currently being probed.

The most stringent probe of baryon number violation is proton decay.
Given the operator of \eqref{Bviolating}, the proton lifetime is
$\tau_p \propto M^4$.  Current bounds on the proton lifetime with
respect to various decay modes are of order $\tau_p \agt 10^{32}\
\text{years}$ and imply $M \agt 10^{15}~\gev$.

Evidence for total lepton number violation is provided by
non-vanishing neutrino masses. The operator of \eqref{Lviolating}
yields a neutrino mass when electroweak symmetry is broken and the
Higgs scalar obtains a vacuum expectation value.  For $\langle h
\rangle \sim m_W$ and $m_{\nu}^2 \sim \Delta m_{\text{atm}}^2 \sim
10^{-3}~\ev^2$, the scale of $L$ violation seen in neutrino
oscillation experiments is $M \sim 10^{14}~\gev$.

Neutrino oscillations also provide evidence for LFV at high scales,
but do not preclude much larger $L_i$-violating, but $L$-conserving,
effects in the charged lepton sector.  For charged LFV, there are a
wide variety of probes~\cite{Kuno:1999jp}, including $\mu \to e
\gamma$, $\mu - e$ conversion, $\mu \to eee$, $\tau \to \mu \gamma$,
$Z\to \mu \tau$~\cite{Zdecay}, $K \to \mu e$~\cite{Molzon}, $J/\Psi
\to \tau \mu$~\cite{Zhang:2000ri}, and others.  All of these may be
useful (see below).  Here, let's concentrate on two of the most
promising: $\mu \to e \gamma$, and $\mu- e$ conversion.  To estimate
the reach of current and future experiments, consider a refined
version of the first operator of \eqref{Liviolating}:
\begin{equation}
{\cal L}_{12} = e \frac{g^2}{16 \pi^2} \frac{m_{\mu}}{M_{12}^2}
\; \bar{\mu} \sigma^{\alpha \beta} e \; F_{\alpha \beta} \ .
\label{specific}
\end{equation}
The additional factors reflect an expectation that the new physics
enters at one-loop and is also suppressed by a chirality insertion
proportional to $m_{\mu}$.  With this parametrization, the rates for
$\mu \to e \gamma$ and $\mu -e$ conversion are
\begin{eqnarray}
\frac{B(\mu \to e \gamma)}{1.2\times 10^{-11}} &=& \left
[ \frac{20~\tev}{M_{12}} \right]^4 \qquad \ \, \text{(MEGA)} 
\nonumber \\
\frac{R(\mu N \to e N)}{6.1\times 10^{-13}} &=& \left
[ \frac{10~\tev}{M_{12}} \right]^4 \qquad \ \, \text{(SINDRUM)}
\nonumber \\
\frac{B(\mu \to e \gamma)}{1\times 10^{-14}} &=& \left
[ \frac{110~\tev}{M_{12}} \right]^4 \qquad 
(\mu e \gamma \text{\ at\ PSI)} \label{bounds} \\
\frac{R(\mu N \to e N)}{5\times 10^{-17}} &=& \left
[ \frac{110~\tev}{M_{12}} \right]^4 \qquad \text{(MECO)} 
\nonumber \\
\frac{R(\mu N \to e N)}{1\times 10^{-18}} &=& \left
[ \frac{280~\tev}{M_{12}} \right]^4 \qquad \text{(PRISM)} \ .
 \nonumber
\end{eqnarray}
In the first two lines the rates are normalized to the best current
bounds from the MEGA~\cite{Brooks:1999pu} and SINDRUM~\cite{SINDRUM}
collaborations.  The last three are normalized to the expected
sensitivities of the $\mu^+ \to e^+ \gamma$ experiment at
PSI~\cite{PSI}, the MECO experiment at Brookhaven~\cite{MECO}, and the
proposed KEK/JAERI joint project, PRISM~\cite{PRISM}. Here $R(\mu N
\to e N) \equiv \Gamma(\mu N \to e N) / \Gamma ( \mu N \to \
\text{capture})$, and I have assumed $R(\mu N \to e N) \approx B(\mu
\to e \gamma) / 300$.  The latter relation depends on the target
nucleus $N$~\cite{nuclei}, but this approximation is adequate for our
purposes. More importantly, I have assumed that the new physics enters
solely through the operator of \eqref{specific}. The sensitivity of
$\mu - e$ conversion relative to $\mu \to e \gamma$ may be very much
greater in other cases.

{}From \eqref{bounds}, two conclusions may be drawn.  First,
experiments underway will extend the sensitivity in mass scale probed
by roughly one order of magnitude.  This is remarkable, given that the
rates scale as $1/M^4$.  Second, the near future sensitivity will be
well above the weak scale.  Many exotic particles may mediate
LFV. These include extra families, $Z'$ bosons, and extra scalars.
All of these particles may exist at or above the weak scale, and so
provide motivation for LFV experiments~\cite{Marciano}.

Particularly compelling, however, is that LFV experiments will soon
stringently probe the weak scale.  Current theories of electroweak
symmetry breaking, which are necessarily tied to the weak scale, are
therefore of great relevance for LFV experiments.  I will discuss some
of these possibilities in the following section. Note that in this
analysis, I have already included factors for loop and chirality
suppression.  In specific models these may be supplemented by
additional suppressions from, for example, GIM-type cancellations or
small intergenerational mixing angles.  A more detailed analysis runs
counter to the model-independent spirit of this section.  In concrete
examples, however, these factors typically serve only to reduce the
${\cal O}(100~\tev)$ reach to a more typical weak scale value of
${\cal O}(1~\tev)$, and observable LFV rates from weak scale physics
are still predicted.

\section{LFV and Electroweak Symmetry Breaking}
\label{sec:EWSB}

The standard model and two possible frameworks for extending it nicely
illustrate some of the possibilities for global symmetry violation:
\begin{equation}
\begin{array}{ccccc}
\text{Example}  & \quad \text{Standard\ Model} \quad & 
\quad \text{Extra\ Dims} \quad & \quad \text{SUSY}
\quad \rule[-1mm]{0mm}{5mm} \\
M_\text{High} & B, L, L_i &           & B, L \rule[-1mm]{0mm}{5mm} \\
M_\text{Weak} &           & B, L, L_i & L_i \rule[-1mm]{0mm}{5mm}
\end{array}
\label{scenarios}
\end{equation}
Here I have indicated the scales at which the various global
symmetries are generically broken.  $M_\text{Weak} \sim 100~\gev$ is
the scale of electroweak symmetry breaking.  $M_\text{High}$
represents some much higher scale, such as the scale of right-handed
neutrinos $M_N \sim 10^{14}~\gev$, grand unification $M_{\text{GUT}}
\approx 2 \times 10^{16}~\gev$, or gravitational interactions
$M_{\text{Planck}} \sim 10^{19}~\gev$.

As previously noted, in the standard model, even extended to include
neutrino masses, all global symmetries are broken only at some high
scale.  In this case, LFV experiments are far from the sensitivity
required, and the exploration of global symmetry violation will be
confined to neutrino and proton decay experiments.  Note that the
observed neutrino mixings do induce charged LFV at the loop level.
However, for $\Delta m_{\nu}^2 \sim 10^{-3}~\ev^2$, the induced LFV is
$B(l \to l' \gamma) \sim 10^{-48}$, far beyond reach.

There are, however, several reasons to expect the standard model to be
incomplete.  Prominent among these is the gauge hierarchy problem, but
there are many others, such as the requirement of sufficient
baryogenesis and the necessity of additional non-baryonic dark matter.
In addition, if current indications for a Higgs boson with mass $m_h
\approx 115~\gev$ are valid, the standard model vacuum will be
destabilized at energy scales $\sim 10^6~\gev$, indicating that some
new physics must enter below this scale~\cite{Sher:1989mj}.

Extra dimensions have been suggested as one way to address the gauge
hierarchy problem~\cite{Antoniadis:1998ig}. With extra dimensions, the
gravitational force law is modified at distances small compared to the
size of the extra dimensions. For particular choices of the number and
size of the extra dimensions, the fundamental gravitational scale may
be lowered to near the weak scale, translating the gauge hierarchy
problem into a problem of hierarchies in length scales.

In such scenarios, one typically expects all global symmetries to be
broken at $M_{\text{grav}} \sim M_\text{Weak}$, leading to, among
other things, catastrophic proton decay. The challenge of finding an
elegant and compelling solution to this difficulty, particularly one
consistent with baryogenesis, is one of the important problems in this
scenario.

It is possible, however, to assume that proton stability is achieved
by some mechanism and consider the possibility of LFV at low scales.
This approach has been taken by a number of authors~\cite{extradims}.
They find that in certain scenarios constraints from LFV are among the
most stringent.  Large rates at future experiments may therefore be
expected in models with low scale gravity, especially if the gravity
scale is not far above the weak scale.

In supersymmetry, a hierarchy between $B$ and $L$ violation on the one
hand and $L_i$ flavor violation on the other arises as an immediate
consequence of the basic motivations for supersymmetry.  As is
well-known, two of the most important phenomenological motivations for
supersymmetry are the gauge hierarchy problem and the existence of
dark matter.  Supersymmetry stabilizes the gauge hierarchy by
introducing at the weak scale a scalar superpartner $\tilde{f}$ for
every standard model fermion $f$ (and also a fermionic superpartner
for every standard model boson).  These new states introduce many
additional renormalizable and gauge-invariant interactions.  In the
notation of superfields, the terms analogous to the standard model's
Yukawa terms of \eqref{Yukawa} are given by the superpotential
\begin{eqnarray}
W &=& y_e H_d L E + y_d H_d Q D + y_u H_u Q U \nonumber \\ 
&& + \lambda L L E + \lambda' L Q D + \lambda'' U D D \ .
\label{superpotential}
\end{eqnarray}
Here generational labels are suppressed, and each superfield contains
both scalar and fermion fields: $F \supset (f, \tilde{f})$.
Interactions are formed by choosing any combination of two fermions
and one scalar from any term.

It is not difficult to see that these new interactions destroy the
beautiful properties of the standard model with respect to global
symmetries.  In particular, the terms in the second line of
\eqref{superpotential} violate both $B$ and $L$.  However, these terms
also allow all superpartners to decay, destroying the possibility of
supersymmetric dark matter.  This important virtue may be preserved by
requiring $R$-parity conservation, where $R \equiv
(-1)^{B+L+2S}$. This eliminates all interactions in the second line of
\eqref{superpotential}, and so $B$ and $L$ are again broken only at a
high scale.  As we will see below, however, the remaining
supersymmetric interactions still allow $L_i$-violating processes at
the weak scale, leading to the hierarchy displayed in
\eqref{scenarios}.

\section{LFV in Supersymmetry}
\label{sec:supersymmetry}

Because LFV experiments will probe weak scale effects, new physics
that is intimately tied to the weak scale is of special importance.  I
now consider several examples in supersymmetry, which provides a
concrete and quantitative framework for evaluating the prospects for
LFV.

Supersymmetric theories are specified by the interactions of
\eqref{superpotential}, and a number of supersymmetry breaking terms.
Of the latter, the most important for our present purposes are slepton
masses
\begin{equation}
m^2{}^{LL}_{ij} \tilde{l}_i^* \tilde{l}_j +
m^2{}^{RR}_{ij} \tilde{e}_i^* \tilde{e}_j \ ,
\label{slepton}
\end{equation}
where the $m^2$ are {\em a priori} arbitrary Hermitian matrices, and
trilinear terms
\begin{equation}
{A_e}_{ij} h_d \tilde{l}_i^* \tilde{e}_j \ ,
\label{Aterm}
\end{equation}
which, after electroweak symmetry breaking, couple left- and
right-handed charged sleptons through the mass matrix $m^2{}^{LR}_{ij}
\equiv {A_e}_{ij} \langle h_d \rangle$.

These terms lead to charged LFV.  To see this, it is convenient to
choose the basis in which the standard model leptons are rotated to
diagonalize the lepton Yukawa coupling, and the sleptons are rotated
to preserve flavor-diagonal gaugino couplings.  No additional flavor
rotations are then available to diagonalize the terms of
Eqs.~(\ref{slepton}) and (\ref{Aterm}), and the off-diagonal elements
in the slepton mass matrices mediate charged LFV.  Define parameters
$\delta^{MN}_{ij} \equiv m^2{}^{MN}_{ij} / \tilde{m}^2$, where
$\tilde{m}^2$ is the representative slepton mass scale, $M, N = L, R$,
and $i,j$ are generational indices.  The branching ratios for $l \to
l' \gamma$, normalized to current
bounds~\cite{Brooks:1999pu,taudecay}, constrain these
parameters\footnote{More precisely, they constrain
$(\delta^{MN}_{ij})_{\rm eff} \sim {\rm max} \left[ \delta^{MN}_{ij},\
\delta^{MP}_{ik} \delta^{PN}_{kj},\ \ldots,\ ( i \leftrightarrow j)
\right]$.} through~\cite{Gabbiani:1996hi}
\begin{eqnarray}
\frac{B(\mu\to e\gamma)}{1.2\times10^{-11}} &\sim& 
{\rm max}\left[ \left(\frac{\delta^{LL,RR}_{12}}
{2.0\times10^{-3}}\right)^2,
                \left(\frac{\delta^{LR}_{12}}
{6.9\times10^{-7}}\right)^2\right]
\left[\frac{100~\gev}{\tilde m}\right]^4 \nonumber \\
\frac{B(\tau\to e\gamma)}{2.7\times10^{-6}} &\sim& 
{\rm max}\left[ \left(\frac{\delta^{LL,RR}_{13}}
{2.2}\right)^2,
                \left(\frac{\delta^{LR}_{13}}
{1.3\times10^{-2}}\right)^2\right]
\left[\frac{100~\gev}{\tilde m}\right]^4           \\
\frac{B(\tau\to \mu\gamma)}{1.1\times10^{-6}} &\sim& 
{\rm max}\left[ \left(\frac{\delta^{LL,RR}_{23}}
{1.4}\right)^2,
                \left(\frac{\delta^{LR}_{23}}
{8.3\times10^{-3}}\right)^2\right]
\left[\frac{100~\gev}{\tilde m}\right]^4 \ . \nonumber 
\end{eqnarray}
Here I have assumed the lightest neutralino to be photino-like with
mass $m^2_{\tilde\gamma}/ \tilde m^2=0.3$; the bounds are fairly
insensitive to this ratio.

The bounds from all three decays are non-trivial, but are especially
stringent for transitions between the first and second generations.
Clearly, arbitrary slepton mass matrices are forbidden.  This is a
statement of the (leptonic) supersymmetric flavor problem.  For
supersymmetry to be viable, some structure in the supersymmetry
breaking masses must be present.  I now turn to a variety of
possibilities and analyze the LFV consequences of each.

\subsection{SU(5)}
\label{sec:su5}

The unification of gauge couplings is a strong motivation to consider
supersymmetric grand unified theories.  In the case of SU(5), the
fields of each generation are contained in two multiplets, the {\bf
10} and $\bar{\bf 5}$, while the up- and down-type Higgs doublets are
contained in the ${\bf 5}_H$ and $\bar{{\bf 5}}_H$ multiplets,
respectively.  The Yukawa couplings are given by the superpotential
\begin{equation}
W_{\text{SU(5)}} = {y_u}_{ij} {\bf 10}_i {\bf 10}_j {\bf 5}_H + 
{y_d}_{ij} \bar{{\bf 5}}_i {\bf 10}_j \bar{{\bf 5}}_H \ .
\end{equation}

What is the form of the scalar masses?  A simple and conservative
assumption is that they are $m_0^2 \widetilde{{\bf 10}}_i^*
\widetilde{{\bf 10}}_i + m_0^2 \widetilde{\bar{{\bf 5}}}_i^*
\widetilde{\bar{{\bf 5}}}_i$, universal, and therefore
flavor-conserving, at $M_{\text{Planck}}$.  However, to determine the
physical consequences, we must evolve these to low energy scales.  In
the renormalization group (RG) evolution above the GUT scale, grand
unified interactions will generate off-diagonal slepton
masses\cite{Barbieri:1994pv}.  This is easily seen by noting that
$y_d$ may be diagonalized with rotations on the {\bf 10} and
$\bar{{\bf 5}}$ fields, but there is then no remaining freedom to
diagonalize $y_u$.  RG evolution will then generate mixing of the {\bf
10} representations, and since $\tilde{e}_R, \tilde{\mu}_R,
\tilde{\tau}_R \in {\bf 10}$, non-vanishing $\delta^{RR}_{ij}$ are
generated.  Detailed calculations show that for some parameter
choices, $\mu^-_L \to e^-_R \gamma$ is already near experimental
limits, and well within reach of future
experiments~\cite{GUTpapers}.  Note that the left-handed sleptons
are not significantly mixed, and so the branching ratio for $\mu^-_R
\to e^-_L \gamma$ is negligible.

\subsection{Right-handed Neutrinos}
\label{sec:right-handed}

A similar analysis may be applied to supersymmetric theories with
right-handed neutrinos $N$.  The leptonic superpotential is
\begin{equation}
W_N = {y_e}_{ij} H_d L_i E_j + {y_{\nu}}_{ij} H_u L_i N_j +
{m_N}_{ij} N_i N_j \ .
\end{equation}
Assume again, conservatively, that the slepton masses are universal at
the Planck scale.  Rotations of the $L$, $E$, and $N$ multiplets may
diagonalize $y_e$ and $m_N$, but then off-diagonal entries in
$y_{\nu}$ remain.  These will generate mixing in the $L$
representations through RG effects above the right-handed neutrino
mass scale, and since $\tilde{e}_L, \tilde{\mu}_L, \tilde{\tau}_L \in
L$, non-vanishing parameters $\delta^{LL}_{ij}$ are generated, which
mediate $\mu^-_R \to e^-_L \gamma$.  These are again within reach of
future experimental sensitivities~\cite{Hisano:1999fj}.  Note that, in
contrast to the case of SU(5), here no $\delta^{RR}_{ij}$ mixing is
generated, and so $\Gamma( \mu^-_L \to e^-_R \gamma) \approx 0$.

\subsection{Flavor Symmetries}
\label{sec:flavor}

In the previous two examples, the Yukawa couplings were taken to have
the hierarchical form required by experiment, and the scalar masses
were assumed to have a simple universal form, presumably following
from some unspecified supersymmetry breaking mechanism.  An orthogonal
approach is try to understand both Yukawa couplings and scalar mass
textures in terms of a single flavor symmetry.  In this approach, one
must first find flavor symmetries that reproduce all of the known
lepton masses and neutrino parameters.  Then, since these symmetries
determine also the superpartner mass matrices, unambiguous predictions
for charged LFV follow.

The existing data may be summarized as follows: letting $\lambda \sim
0.2$ be the small flavor breaking parameter, the charged lepton masses
satisfy
\begin{equation}
m_\tau/\langle h_d \rangle \sim \lambda^3 - 1, \quad
m_\mu/m_\tau \sim \lambda^2, \quad
m_e/m_\mu \sim \lambda^3 \ ,
\end{equation}
and the neutrino data imply
\begin{eqnarray}
&&\sin \theta_{23} \sim 1, \quad
\sin \theta_{13} \alt \lambda, \quad
\sin \theta_{12} \sim \left\{ \begin{array}{ll}
1\ \ & \text{MSW(LA), VO} \\ \lambda^2 & \text{MSW(SA)} 
\end{array} \right. \ , \nonumber \\
&&\frac{\Delta {m_{\nu}^2}_{12}}{\Delta {m_{\nu}^2}_{23}}
\sim\left\{ \begin{array}{ll}
\lambda^2-\lambda^4 \ \ & \text{MSW} \\ \lambda^8-\lambda^{12} 
& \text{VO} \end{array} \right. \ ,
\label{nudata}
\end{eqnarray}
where MSW(LA,SA) and VO denote the various solar neutrino solutions.

These data cannot be reproduced by the simplest possibility, a single
U(1) symmetry.  In this case, the fermion mass matrices are determined
by charge assignments $Q$ as
\begin{equation}
{m_e}_{ij} \sim \langle h_d \rangle
\lambda^{Q(\bar{l}_i)+Q(e_j)+Q(h_d)} \ , \ \
{m_\nu}_{ij} \sim \frac{\langle h_u \rangle^2}{M_N} 
\lambda^{Q(l_i)+Q(l_j)+2Q(h_u)} \ , 
\end{equation}
with similar expressions for the scalar masses.  It is not difficult
to show that these textures predict that highly mixed neutrinos have
similar masses, contradicting the most straightforward interpretation
of \eqref{nudata}.

However, the observed data may be explained by, for example, $Z_n
\times$ U(1) symmetries, two breaking parameters with opposite
charges, or holomorphic zeros~\cite{abelian}.  In these models, the
above-mentioned difficulty is almost always circumvented as follows:
the $m_\nu$ matrix produces hierarchical neutrino masses, and the
$m_e$ matrix generates large neutrino mixing by requiring that the
gauge and mass eigenstates of the $l$ representations are related by
large rotations.  However, because the lepton doublets contains
charged leptons in addition to neutrinos, these rotation also
generates large misalignments between the charged leptons and
sleptons, producing large charged LFV rates~\cite{Feng:2000wt}.

It is also possible for the $m_\nu$ matrix to have a very special form
such that it singlehandedly generates both large mixings and large
hierarchies in neutrino masses.  Such examples have been
constructed~\cite{Feng:2000wt}.  Typically, however, the former
possibility is realized.  Large angle neutrino solutions then imply
large LFV rates well within reach of future $\mu \to e \gamma$ and
$\mu - e$ conversion experiments~\cite{Feng:2000wt}.  Conversely, if
no signal is seen in future experiments, many supersymmetric flavor
models for neutrino masses will be excluded.

\section{What Will We Learn?}
\label{sec:what}

\subsection{Precision LFV}

Given the dramatic improvements expected at future experiments, it is
possible not only that LFV will be discovered, but also that it will
be discovered with large statistics, allowing precision studies.
Comparisons of the different rates of, for example, $\mu \to e
\gamma$, $\mu - e$ conversion, and $\mu \to eee$ will allow us to
disentangle which of the various operators in \eqref{Liviolating} are
contributing~\cite{Tobe}.

The distributions within one mode may also be informative.  With
polarized muons, angular distributions in $\mu \to e \gamma$, and
possibly also in $\mu - e$ conversion, may be used to disentangle the
two chirality components.  This will be a powerful way to distinguish
between different models~\cite{Kuno:1996kv}, as illustrated by the
examples in Secs.~\ref{sec:su5} and \ref{sec:right-handed}.

\subsection{Interplay with Colliders}

If LFV is discovered in the near future, it is quite possible that the
LFV-inducing particles will be produced directly at high energies.
LFV effects from real particle production may then also be seen at
colliders.  In the case of supersymmetry, sleptons produced in one
flavor eigenstate may oscillate to another before decaying, leading to
anomalously large rates for $e \mu +X$ production, for
example~\cite{sleptonLFV}.  In the case of two generation mixing, the
cross section for such processes in the simplest cases is
$\sigma(e\mu) \propto \sin^2 2 \theta_{12}\, (\Delta m_{12}^2)^2 /
[(\Delta m_{12}^2)^2 + 4 \tilde{m}^2 \Gamma^2]$, where $\theta_{12}$
and $\Delta m_{12}^2$ are the mixing angle and mass splitting between
the first two slepton generations, and $\tilde{m}$ and $\Gamma$ are
the average slepton mass and decay width.  The rate for low energy
$\mu-e$ processes has a different dependence on mixing parameters,
with $B(\mu \to e) \propto \sin^2 2 \theta_{12} \left[\Delta m_{12}^2
/ \tilde{m}^2 \right]^2$.  Thus, the combination of high energy and
high precision experiments may allow us to determine both mixing
angles and mass splittings independently, which would be impossible
with only one type of experiment.

\subsection{Null Results}

If no signal is seen, the implications depend on what is discovered at
high energy colliders.  If only a standard model-like Higgs boson is
discovered there, the null LFV results will simply give powerful
constraints on a variety of phenomena.  Of course, in this case, large
questions will remain concerning the physics of electroweak symmetry
breaking.

More likely, some part of the physics of electroweak symmetry breaking
will be uncovered at colliders.  We have seen that, on very general
grounds, theories of electroweak symmetry breaking often predict LFV
effects within reach of the upcoming experiments.  Null results from
these experiments, then, will provide highly non-trivial guides to
understanding these theories.  In the case of supersymmetry, for
example, null results may have far-reaching implications, excluding
many currently attractive supergravity theories and favoring theories
that generate extremely degenerate superpartners at a low scale.

\section*{Acknowledgments}
I thank Y.~Kuno and W.~Molzon for the invitation to participate in
this stimulating conference, and I am grateful to Y.~Nir and Y.~Shadmi
for collaboration in the work described in Sec.~\ref{sec:flavor}.
This work was supported in part by funds provided by the
U.~S.~Department of Energy under cooperative research agreement
DF--FC02--94ER40818.

%\section*{Appendix}


\section*{References}

\begin{thebibliography}{99}

\bibitem{Kuno:1999jp}
For comprehensive reviews, see Y.~Kuno and Y.~Okada,
%``Muon decay and physics beyond the standard model,''
hep-ph/9909265; and references therein.
%%CITATION = HEP-PH 9909265;%%

\bibitem{'tHooft}
%\bibitem{'tHooft:1976up}
G.~'t Hooft,
%``Symmetry breaking through Bell-Jackiw anomalies,''
Phys.\ Rev.\ Lett.\ {\bf 37}, 8 (1976);
%%CITATION = PRLTA,37,8;%%
%\bibitem{'tHooft:1976fv}
%G.~'t Hooft,
%``Computation of the quantum effects due to a four-dimensional 
%pseudoparticle,''
Phys.\ Rev.\ D {\bf 14}, 3432 (1976).
%%CITATION = PHRVA,D14,3432;%%

\bibitem{Appelquist:1975tg}
T.~Appelquist and J.~Carazzone,
%``Infrared Singularities And Massive Fields,''
Phys.\ Rev.\ D {\bf 11}, 2856 (1975).
%%CITATION = PHRVA,D11,2856;%%

\bibitem{Zdecay}
%\bibitem{Illana:2000ic}
See, for example, 
J.~I.~Illana and T.~Riemann,
%``Charged lepton flavour violation from massive neutrinos in Z decays,''
hep-ph/0010193;
%%CITATION = HEP-PH 0010193;%%
%\bibitem{Bi:2000xp}
X.~Bi, Y.~Dai and X.~Qi,
%``Lepton flavor violation in supersymmetric SO(10) grand unified
%models,''
hep-ph/0010270.
%%CITATION = HEP-PH 0010270;%%

\bibitem{Molzon}
W.~Molzon, these proceedings.

\bibitem{Zhang:2000ri}
See, for example, 
X.~Zhang,
%``Probing for new physics in J/psi decays,''
hep-ph/0010105.
%%CITATION = HEP-PH 0010105;%%

\bibitem{Brooks:1999pu}
M.~L.~Brooks {\it et al.}  [MEGA Collaboration],
%``New limit for the family-number non-conserving decay 
%mu+ --> e+ gamma,''
Phys.\ Rev.\ Lett.\ {\bf 83}, 1521 (1999)
[hep-ex/9905013].
%%CITATION = HEP-EX 9905013;%%

\bibitem{SINDRUM}
P.~Wintz, these proceedings.
%P.~Wintz [Sindrum II Collaboration], in {\it Proceedings of Lepton and
%Baryon Number Violation in Particle Physics, Astrophysics and
%Cosmology}, eds. H.~V.~Klapdor-Kleingrothaus and I.~V.~Krivosheina
%(Institute of Physics Publishing, Bristol and Philadelphia, 1999),
%p.~534.

\bibitem{PSI}
J.~Yashima, these proceedings.
% http://meg.psi.ch.

\bibitem{MECO}
J.~Sculli, these proceedings; V.~Tumakov, these proceedings.
%http://meco.ps.uci.edu.

\bibitem{PRISM}
M.~Aoki, these proceedings; Y.~Kuno, these proceedings.
%http://psux1.kek.jp/$\sim$prism.

\bibitem{nuclei}
See, for example, 
A.~Czarnecki, W.~J.~Marciano and K.~Melnikov, hep-ph/9801218; 
%\bibitem{Kim:1999um}
Y.~G.~Kim, P.~Ko, J.~S.~Lee and K.~Y.~Lee,
%``Complete analysis of photino-mediated lepton flavor violations in
%generalized supersymmetric models,''
Phys.\ Rev.\ D {\bf 59}, 055018 (1999)
[hep-ph/9811211];
%%CITATION = HEP-PH 9811211;%%
%\bibitem{Kosmas:1997ec}
T.~S.~Kosmas, A.~Faessler, F.~Simkovic and J.~D.~Vergados,
%``State-by-state calculations for all channels of the exotic (mu-,e-)
% conversion process,''
Phys.\ Rev.\ C {\bf 56}, 526 (1997)
[nucl-th/9704021];
%%CITATION = NUCL-TH 9704021;%%
A.~Czarnecki, these proceedings; 
T.~S.~Kosmas, these proceedings.

\bibitem{Marciano}
W.~Marciano, these proceedings.

\bibitem{Sher:1989mj}
M.~Sher,
%``Electroweak Higgs Potentials And Vacuum Stability,''
Phys.\ Rept.\ {\bf 179}, 273 (1989).
%%CITATION = PRPLC,179,273;%%

\bibitem{Antoniadis:1998ig}
I.~Antoniadis, N.~Arkani-Hamed, S.~Dimopoulos and G.~Dvali,
%``New dimensions at a millimeter to a Fermi and superstrings at a TeV,''
Phys.\ Lett.\ {\bf B436}, 257 (1998)
[hep-ph/9804398].
%%CITATION = HEP-PH 9804398;%%

\bibitem{extradims}
%\bibitem{Faraggi:1999bm}
See, for example, A.~E.~Faraggi and M.~Pospelov,
%``Phenomenological issues in TeV scale gravity with light neutrino  
%masses,''
Phys.\ Lett.\ {\bf B458}, 237 (1999)
[hep-ph/9901299];
%%CITATION = HEP-PH 9901299;%%
%\bibitem{Silagadze:1999hr}
Z.~K.~Silagadze,
%``Lepton-flavor violating decays as probes of quantum gravity?,''
hep-ph/9907328;
%%CITATION = HEP-PH 9907328;%%
%\bibitem{Ioannisian:2000cw}
A.~Ioannisian and A.~Pilaftsis,
%``Cumulative non-decoupling effects of Kaluza-Klein neutrinos in  
%electroweak processes,''
Phys.\ Rev.\ D {\bf 62}, 066001 (2000)
[hep-ph/9907522];
%%CITATION = HEP-PH 9907522;%%
%\bibitem{Kitano:2000wr}
R.~Kitano,
%``Lepton flavor violation in the Randall-Sundrum model with bulk 
%neutrinos,''
Phys.\ Lett.\ {\bf B481}, 39 (2000)
[hep-ph/0002279].
%%CITATION = HEP-PH 0002279;%%

\bibitem{taudecay}
%\bibitem{Edwards:1997te}
K.~W.~Edwards {\it et al.}  [CLEO Collaboration],
%``Search for neutrinoless tau decays: tau --> e gamma and tau --> mu
%gamma,''
Phys.\ Rev.\ D {\bf 55}, 3919 (1997);
%%CITATION = PHRVA,D55,3919;%%
%\bibitem{Ahmed:2000gh}
S.~Ahmed {\it et al.}  [CLEO Collaboration],
%``Update of the search for the neutrinoless decay tau --> mu gamma,''
Phys.\ Rev.\ D {\bf 61}, 071101 (2000)
[hep-ex/9910060].
%%CITATION = HEP-EX 9910060;%%

\bibitem{Gabbiani:1996hi}
F.~Gabbiani, E.~Gabrielli, A.~Masiero and L.~Silvestrini,
%``A complete analysis of FCNC and CP constraints in general SUSY
%extensions of the standard model,''
Nucl.\ Phys.\ {\bf B477}, 321 (1996)
[hep-ph/9604387].
%%CITATION = HEP-PH 9604387;%%

\bibitem{Barbieri:1994pv}
R.~Barbieri and L.~J.~Hall,
%``Signals for supersymmetric unification,''
Phys.\ Lett.\ {\bf B338}, 212 (1994)
[hep-ph/9408406].
%%CITATION = HEP-PH 9408406;%%

\bibitem{GUTpapers}
%\bibitem{Hisano:1997qq}
J.~Hisano, T.~Moroi, K.~Tobe and M.~Yamaguchi,
%``Exact event rates of lepton flavor violating processes in
%supersymmetric SU(5) model,''
Phys.\ Lett.\ {\bf B391}, 341 (1997)
[hep-ph/9605296];
%%CITATION = HEP-PH 9605296;%%
%\bibitem{Hisano:1998cx}
J.~Hisano, D.~Nomura, Y.~Okada, Y.~Shimizu and M.~Tanaka,
%``Enhancement of mu --> e gamma in the supersymmetric SU(5) GUT at
%large  tan(beta),''
Phys.\ Rev.\ D {\bf 58}, 116010 (1998)
[hep-ph/9805367];
%%CITATION = HEP-PH 9805367;%%
Y.~Okada, these proceedings.

\bibitem{Hisano:1999fj}
J.~Hisano and D.~Nomura,
%``Solar and atmospheric neutrino oscillations and lepton flavor
%violation  in supersymmetric models with the right-handed
%neutrinos,''
Phys.\ Rev.\ D {\bf 59}, 116005 (1999)
[hep-ph/9810479];
%%CITATION = HEP-PH 9810479;%%
D.~Nomura, these proceedings.

\bibitem{abelian}
%\bibitem{Grossman:1998jj}
See, for example, Y.~Grossman, Y.~Nir and Y.~Shadmi,
%``Large mixing and large hierarchy between neutrinos with Abelian
%flavor  symmetries,''
JHEP{\bf 9810}, 007 (1998)
[hep-ph/9808355];
%%CITATION = HEP-PH 9808355;%%
%\bibitem{Nir:1999xp}
Y.~Nir and Y.~Shadmi,
%``Testing Abelian flavor symmetries with neutrino parameters,''
JHEP{\bf 9905}, 023 (1999)
[hep-ph/9902293].
%%CITATION = HEP-PH 9902293;%%

\bibitem{Feng:2000wt}
J.~L.~Feng, Y.~Nir and Y.~Shadmi,
%``Neutrino parameters, Abelian flavor symmetries, and charged lepton
%flavor violation,''
Phys.\ Rev.\ D {\bf 61}, 113005 (2000)
[hep-ph/9911370].
%%CITATION = HEP-PH 9911370;%%

\bibitem{Tobe}
%\bibitem{deGouvea:2001cf}
A.~de Gouvea, S.~Lola and K.~Tobe,
%``Lepton flavor violation in supersymmetric models with trilinear
%R-parity violation,''
Phys.\ Rev.\ D {\bf 63}, 035004 (2001)
[hep-ph/0008085];
%%CITATION = HEP-PH 0008085;%%
%\bibitem{Tobe:2000bv}
K.~Tobe,
%``Lepton flavor violation in SUSY models with and without R-parity,''
hep-ph/0008075;
%%CITATION = HEP-PH 0008075;%%
K.~Tobe, these proceedings.

\bibitem{Kuno:1996kv}
Y.~Kuno and Y.~Okada,
%``$\mu\to e\gamma$ Search with Polarized Muons,''
Phys.\ Rev.\ Lett.\ {\bf 77}, 434 (1996)
[hep-ph/9604296].
%%CITATION = HEP-PH 9604296;%%

\bibitem{sleptonLFV}
%\bibitem{Arkani-Hamed:1996au}
N.~Arkani-Hamed, H.~Cheng, J.~L.~Feng and L.~J.~Hall,
%``Probing Lepton Flavor Violation at Future Colliders,''
Phys.\ Rev.\ Lett.\ {\bf 77}, 1937 (1996)
[hep-ph/9603431];
%%CITATION = HEP-PH 9603431;%%
%\bibitem{Arkani-Hamed:1997km}
\ibid,
%``CP violation from slepton oscillations at the LHC and NLC,''
Nucl.\ Phys.\ {\bf B505}, 3 (1997)
[hep-ph/9704205];
%%CITATION = HEP-PH 9704205;%%
M.~Tanaka, these proceedings.

\end{thebibliography}
\end{document}